\documentclass[journal=jctcce,manuscript=article]{achemso}

\usepackage[version=3]{mhchem} 
\usepackage{graphicx}
\usepackage{dcolumn}
\usepackage{bm}
\usepackage[font=small]{caption}
\usepackage{subcaption}
\usepackage[np]{numprint}
\nprounddigits{2}
\npdecimalsign{.}
\usepackage{xcolor}
\usepackage[normalem]{ulem}
\usepackage{amsmath}
\usepackage{amssymb}
\usepackage{soul}

\newcommand*{\rv}{\mathbf{r}}
\newcommand*{\xv}{\mathbf{x}}

\setstcolor{red}

\newif\ifrevision
\revisiontrue 
\ifrevision
\newcommand{\revstrike}[1]{\st{#1}}
\else

\newcommand{\revstrike}[1]{}
\fi

\setkeys{acs}{maxauthors=10}
\setkeys{acs}{etalmode=truncate}

\SectionNumbersOn

\author{Annina Z. Lieberherr}
\affiliation{Department of Chemistry, University of Oxford, Physical and Theoretical Chemistry Laboratory, South Parks Road, Oxford OX1 3QZ, United Kingdom}
 \email{annina.lieberherr@chem.ox.ac.uk}

\author{Paola Gori-Giorgi}
\altaffiliation{Microsoft Research AI for Science, Evert van de Beekstraat 354, 1118CZ Schiphol,
The Netherlands}
\affiliation{Department of Chemistry and Pharmaceutical Sciences, Amsterdam Institute of Molecular and Life Sciences (AIMMS), Faculty of Science, Vrije Universiteit Amsterdam, De Boelelaan 1083, 1081HV Amsterdam, The Netherlands}

\author{Klaas J. H. Giesbertz}
\affiliation{Department of Chemistry and Pharmaceutical Sciences, Amsterdam Institute of Molecular and Life Sciences (AIMMS), Faculty of Science, Vrije Universiteit Amsterdam, De Boelelaan 1083, 1081HV Amsterdam, The Netherlands}

\title[Optimal transport distances to characterise electronic excitations]{Optimal transport distances to characterise electronic excitations}

\abbreviations{IR,NMR,UV}

\begin{document}




\begin{abstract}
Understanding the character of electronic excitations is important in computational and reaction mechanistic studies, but their classification from simulations remains an open problem. Distances based on optimal transport have proven very useful in a plethora of classification problems and seem therefore a natural tool to try to tackle this challenge.
We propose and investigate a new diagnostic $\Theta$ based on the Sinkhorn divergence from optimal transport.
We evaluate a $k$-NN classification algorithm on $\Theta$, the popular $\Lambda$ diagnostic as well as their combination, and assess their performance in labelling excitations, finding that
(i) The combination only slightly improves the classification, (ii) Rydberg excitations are not separated well in any setting,
and (iii) $\Theta$ breaks down for charge transfer in small molecules.
We then define a length scale-normalised version of $\Theta$ and show that the result correlates closely with $\Lambda$ for results obtained with Gaussian basis functions. Finally, we discuss the orbital-dependence of our approach and explore an orbital-independent version. Using an optimised combination of the optimal transport and overlap diagnostics together with a different metric is in our opinion the most promising for future classification studies.
\end{abstract}


\section{Introduction}
\label{sec:introduction}


Electronic excitations drive various processes, for example in light harvesting or solar cells, which are important in the exploration of alternative energy sources.\cite{Scho2017,Hagf2010}
There is consequently significant theoretical interest in calculating excited states and their energies, which could then be used to drive non-adiabatic dynamics and design novel materials.
However, excited state calculations are a computationally expensive problem, and various approximations have been introduced to obtain solutions at lower costs.
Similarly to the ground state problem, methods based on density functional theory (DFT) have become highly popular as they come at a relatively low computational cost but have a solid theoretical foundation, except for approximations that have to be made to the density functional.\cite{Canc2023}
Despite decades of effort, there is no `one-fits-all' density functional.\cite{Medv2017}
In particular for electronic excitations in time-dependent DFT (TDDFT),\cite{Ullr2006} where transition energies are calculated within the linear response setting, it has become clear that a functional's performance can depend heavily on the character of the excitation.\cite{Mait2017}

%
This is one of the main motivations behind exploring diagnostics that split TDDFT calculated excitations into different types,\cite{Peac2008,Mewe2015} of which there are three.
First, Rydberg excitations to energetically high lying and diffuse Rydberg states.
Secondly, charge transfer (CT) excitations from a donor to an acceptor which is spatially separated from the donor, either within the same molecule (intramolecular) or in different molecules (intermolecular).
CT excitation can carry especially large errors in TDDFT calculations.\cite{Mait2017}
The third group are local excitations, which encompass excitations that fit neither of the other two categories.
With a classifier one can decide a posteriori whether the chosen density functional was suitable and if not, repeat the calculation with a higher-level functional.
However, such a classification might prove useful beyond picking the right functional.
Assigning a label to a new excitation can be a very tedious process, during which one relies on chemical intuition or has to inspect wavefunctions and/or localised orbitals visually.
With a quantitative method to decide on the character of the excitation, this step could be extremely simplified.\cite{Hiro2017}

%
To this end, Peach \textit{et al.} took advantage of the decomposition of an electronic excitation into single orbital transitions within the TDDFT framework and studied a weighted average of the overlap between initial and final orbitals.\cite{Peac2008}
While they did not establish a classification of excitation types, they showed a correlation between their overlap diagnostic and the associated error in excitation energy.
The same trends have been observed in other systems and the overlap diagnostic has been widely applied since its introduction.\cite{Dev2012,Lean2012,Korn2011}
In a similar spirit Guido \textit{et al.} studied the difference between the electron's centroid in the initial and final orbitals within a single orbital transition.\cite{Guid2013}
The centroid difference is better able to differentiate between CT and valence excitations than the overlap diagnostic, although they stress that the best results were obtained when using them in combination.

An alternative interpretation of electronic excitations is via the formation of an electron-hole pair (the exciton).
The resulting exciton formalism opens the door for a multitude of other descriptors like the distance between electron and hole, $d_\mathrm{he}$, the electron/hole sizes, $\sigma_e/\sigma_h$ respectively, and the size of the exciton $d_\mathrm{exc}$. All have been used to characterise excited states\cite{Mewe2015,Mewe2019} and the overlap diagnostic has been combined with $d_{he}$ and $d_\mathrm{exc}$ for classification.\cite{Hiro2017}

Going back to a density picture in the TDDFT framework, Moore \textit{et al.} consider the modulus of the change in electron density between the initial and final orbitals, which they argue is better suited for a classification.\cite{Moor2015}
However, this diagnostic suffers from a similar problem as the overlap:
If the sets of points where the initial and final orbitals are nonzero, which will also be referred to as their 'supports', become disjoint, there is a plateau value for both of them and they are agnostic to any variations in the orbitals beyond that point.

In this work, we seek to improve the classification of excitations by considering optimal transport properties of electron densities, which promise to avoid the plateau problem.
The article is structured as follows. After laying out the necessary theory on TDDFT and optimal transport in Sec.~\ref{sec:theory}, we cover the computational details in Sec.~\ref{sec:computational}.
In Sec.~\ref{sec:results}, we present the results, both confirming previous studies on the overlap diagnostic and new results using descriptors from optimal transport.
Finally, we summarise our findings in Sec.~\ref{sec:conclusion} and propose future directions.

\section{Theory}
\label{sec:theory}

\subsection{TDDFT and the $\boldsymbol{\Lambda}$ diagnostic}
\label{subsec:tddft}

The Kohn--Sham (KS) ground state for an $N$-electron system is a Slater determinant
\begin{equation}
	\Psi_{s}(\xv_1, \dotsc, \xv_N) = \mathcal{A} \prod_{i=1}^N \varphi_i(\xv_i), \label{eq:slaterdeterminant}
\end{equation}
where $\mathcal{A}$ anti-symmetrises and normalises the wave function, $\xv_i$ is the space-spin coordinate of the $i$-th electron and $\varphi_i$ is the $i$-th KS orbital.\cite{Canc2023}
The orbitals are solutions to the (time-independent) KS equations,\cite{Kohn1965}
\begin{equation}
    \left[ -\frac{1}{2} \Delta + v_\mathrm{s}(\rv) \right] \varphi_i(\xv) = \varepsilon_i \varphi_i(\xv) .  \label{eq:ksequations}
\end{equation}
The ground state is composed of the KS orbitals $\varphi_i$ with the lowest energies $\varepsilon_i$ (the occupied orbitals).
Excited states within linear-response TDDFT are linear combinations of Slater determinants with single orbital excitations into previously unoccupied (or virtual) orbitals $\varphi_a$.\cite{Ullr2006}
The excitation energy and the contribution of a single orbital transition can be determined in the linear response picture by solving the Casida equations, \cite{Casi1995}
\begin{equation}
    \begin{pmatrix}
    A & B \\ B & A
    \end{pmatrix}
    \begin{pmatrix}
        X \\ Y
    \end{pmatrix}
    =
    \Omega \begin{pmatrix}
        -1 & 0 \\ 0 & 1
    \end{pmatrix}
    \begin{pmatrix}
        X \\ Y
    \end{pmatrix}. \label{eq:casida}
\end{equation}
The eigenvalues $\Omega$ are the excitation energies and the eigenvectors $X, Y$ contain the excitation and de-excitation amplitudes, respectively.
Although equation~\eqref{eq:casida} looks simple, the entries of $A$ and $B$ depend on the exact ground state of the system as well as on the exact form of the exchange correlation functional.
Neither of these is currently attainable, but there are various approximations to the exchange-correlation functional with which excitation energies can be obtained.\cite{Toul2022}
However, TDDFT excitation energies can vary greatly in accuracy between different excitation characters and between different functionals.\cite{Mait2017}

It is generally difficult to assign one of the labels ``CT", ``local" or ``Rydberg" to a given electronic excitation.
Peach \textit{et al.} investigated energy errors in electronic excitations based on the overlap of the single orbital transitions, \cite{Peac2008}
\begin{equation}
    \Lambda = \frac{\sum_{i,a} c_{ia} \left\langle\; |\varphi_a|\; \middle|\; | \varphi_i| \;\right\rangle }{\sum_{i,a} c_{ia}}, \label{eq:def-lambda}
\end{equation}
where $c_{ia} = X_{ia}^2 + Y_{ia}^2$ is the contribution of the excitaiton $\varphi_i \rightarrow \varphi_a$ and $\left\langle\; |\varphi_a|\; \middle|\; | \varphi_i| \;\right\rangle$ is the overlap of the (modulus of the) orbitals. They studied a set of eleven molecules with the PBE, B3LYP and CAM-B3LYP functionals and evaluated $\Lambda$ for a total of 59 excitations.
While $\Lambda$ gives some insight into the expected error magnitude for different DFT functionals,
it is not able to distinguish between the different excitation types without prior knowledge: Though $\Lambda$ values of local and Rydberg excitations are well separated, CT excitations fall across almost the whole range of $\Lambda$.
Hence, one cannot make conclusions about the excitation type based on $\Lambda$ alone.
Optimal transport might be able to help.

\subsection{An optimal transport diagnostic}
\label{subsec:otdiagnostic}

At the heart of optimal transport lies the problem of finding a plan to transport a source probability density into a target probability density.\cite{Peyr2019}
Imagine a pile of soil next to a hole in the ground. With a wheelbarrow, one can -- with some physical effort -- carry the soil over and fill the hole. When the hole is entirely filled, the total work that was necessary is the invested `cost'. For this process, the optimal transport problem would be to plan the wheelbarrow transport such as to minimise the work.
As we will be looking at electronic orbitals $\varphi_{i,a}$, it is only natural to use the probability densities $\rho_{i,a} = \lvert\varphi_{i,a}\rvert^2$.
The (entropically regularised) optimal transport problem is\cite{Cutu2013}
\begin{align}
    \mathcal{W}_{2,\varepsilon}^2(\rho_i, \rho_a) 
    = \min_{\pi \in \Pi(\rho_i, \rho_a)}   \int_{\mathbb{R}^3\times \mathbb{R}^3} &  c(\boldsymbol{r},\boldsymbol{r}') \pi(\boldsymbol{r},\boldsymbol{r}') \mathrm{d} \boldsymbol{r} \mathrm{d} \boldsymbol{r}'  \nonumber\\
    &+ \varepsilon \mathrm{KL}\left(\pi \middle| \rho_i\otimes \rho_a\right), \label{eq:def-ot-entropic}
\end{align}
where $c$ is called a cost function, $\Pi(\rho_i,\rho_a)$ is the set of joint probability densities $\pi$ with marginals $\rho_i$ and $\rho_a$,
\begin{equation}
    \int_{\mathbb{R}^3} \pi(\boldsymbol{r},\boldsymbol{r}') \mathrm{d}\boldsymbol{r}' = \rho_i(\boldsymbol{r}), \hspace{.2cm}  \int_{\mathbb{R}^3} \pi(\boldsymbol{r},\boldsymbol{r}') \mathrm{d}\boldsymbol{r} = \rho_a(\boldsymbol{r}'), \label{eq:def-marginals}
\end{equation}
and we use the regularised optimal transport problem problem with the Kullback--Leibler divergence $\mathrm{KL}\left(\pi \middle| \rho_i \rho_a\right)$
\begin{align}
    \mathrm{KL}(\pi | \nu) =& \int_{\mathbb{R}^3 \times \mathbb{R}^3} \pi( \boldsymbol{r}, \boldsymbol{r}')  \log \frac{\pi( \boldsymbol{r}, \boldsymbol{r}')}{\nu( \boldsymbol{r}, \boldsymbol{r}')}  
    \mathrm{d}\boldsymbol{r} \mathrm{d}\boldsymbol{r}'  \nonumber\\
    &+ \int_{\mathbb{R}^3 \times \mathbb{R}^3} 
     \left(\nu( \boldsymbol{r}, \boldsymbol{r}') - \pi( \boldsymbol{r}, \boldsymbol{r}') \right) \mathrm{d}\boldsymbol{r} \mathrm{d}\boldsymbol{r}'
\end{align}
for computational efficiency.\cite{Cutu2013}
Because of the entropic regularisation, Eq.~\eqref{eq:def-ot-entropic} can unituitively be non-zero even if the orbitals $\varphi_i$ and $\varphi_a$ are identical, which can be fixed by defining the Sinkhorn divergence,\cite{Ramd2017}
\begin{align}
    S(\rho_i,\rho_a) =& \mathcal{W}_{2,\varepsilon}^2(\rho_i, \rho_a) - \frac{1}{2} \left[ \mathcal{W}_{2,\varepsilon}^2(\rho_i, \rho_i) + \mathcal{W}_{2,\varepsilon}^2(\rho_a, \rho_a) \right], \label{eq:def-sinkhorndivergence}
\end{align}
which will therefore be used from here on.
The Sinkhorn divergence $S$ has the same units as the cost function $c$. Depending on the application, different cost functions are appropriate. Here, we will always use the squared Euclidean cost, $c(\boldsymbol{r}, \boldsymbol{r}') = \lVert \boldsymbol{r} - \boldsymbol{r}' \rVert^2$, for which $S$ has units of a length squared.
Note that $\pi$ has units of length$^{-6}$, since it is a joint probability density of two position vectors. $\varepsilon$ also needs to have the same units as $c$, since the Kullback--Leibler divergence is dimensionless.

\begin{figure}[h]
    \centering
    \includegraphics{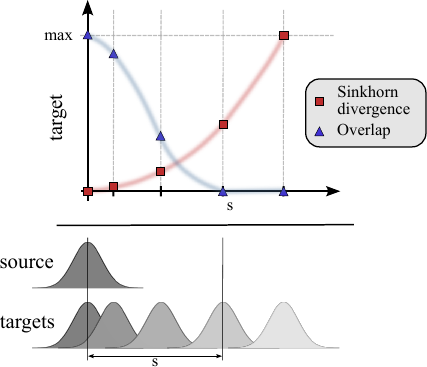}
    \caption{(Qualitative) demonstration of overlap measurements and the Sinkhorn divergence when studying translations, here for Gaussian source and target densities. }
    \label{fig:compmetrics}
\end{figure}

Fig.~\ref{fig:compmetrics} motivates the use of the Sinkhorn divergence to study electronic excitations:
Consider the case where the target density ($\rho_a$ in Eq.~\eqref{eq:def-ot-entropic}) is a translation of the source density $\rho_i$.
This could be a crude model of CT with an increasing number of linker fragments between donor and acceptor, for which Mewes and Dreuw have already found a linear dependence of exciton-based diagnostics.\cite{Mewe2019}
As soon as the supports of source and target density become disjoint, the overlap is zero and stays zero for any larger translations.
It is in that sense blind to translations above some threshold, and $\Lambda$ will tend towards zero.
The optimal transport derived quantities are expected to behave very differently. If the regularisation parameter $\varepsilon$ goes to 0, the squared Wasserstein distance $\mathcal{W}_{2,0}^2$ would be quadratic in the displacement.
While the Sinkhorn divergence $S$ overestimates the Wasserstein distance, it should behave similarly for small enough $\varepsilon$.
The Sinkhorn divergence's sensitivity for translations motivates us to propose a new diagnostic,
\begin{equation}
    \Theta = \frac{\sum_{i,a} c_{ia} S(\rho_i,\rho_a)}{\sum_{i,a} c_{ia}}, \label{eq:def-theta}
\end{equation}
with the single orbital contribution $c_{ia}$ as in Eq.~\eqref{eq:def-lambda}.

\section{Computational details}
\label{sec:computational}

We calculated $\Lambda$ and $\Theta$ for the same set of molecules as studied by Peach \textit{et al.}, because their data set includes many examples of the three different excitation types.\cite{Peac2008}
The set includes the four small molecules N$_2$, CO, HCl and formaldehyde, five larger molecules (dipeptide, $\beta$-dipeptide, tripeptide, N-phenylpyrrole (PP) and 4-(N,N-dimethylamino)benzonitrile (DMABN)) and conjugated polymer systems: acenes (1-5 monomers) and polyacetylene (PA) oligomers (2-5 monomers), all of which are shown in Fig. S1. 
Calculating $\Lambda$ allows us to verify our data before studying the new diagnostic $\Theta$.
Electronic excitation energies were calculated with Turbomole,\cite{Bala2020} using the $d$-aug-cc-pVTZ basis set for N$_2$, H$_2$CO and CO and the cc-pVTZ basis set for the remaining molecules.\cite{Dunn1989,Kend1992,Woon1994,Schu2007,Prit2019}
The SCF convergence threshold was 10$^{-7}$ and the multiple grid \texttt{m3} was used for the DFT calculation.\cite{Eich1997}
At the end of a converged TDDFT calculation, Turbomole prints a list of single orbital contributions $c_{ia}$ (for $c_{ia} > 0.05$) and the involved orbitals, which can be obtained on a grid.
The grid data is then used to calculate $\Lambda$ by numerical quadrature and $\Theta$ with the \texttt{geomloss} package (version \texttt{0.2.5}).\cite{Feyd2019}
In order to calculate Sinkhorn divergences, the orbital grids have to be chosen large enough to contain all electronic density (in practice, we chose the grids big enough to contain at least 99\% of the density).
We use an equidistant grid with spacing of 0.8 Bohr for all molecules.

In order to gain some quantitative understanding of the degree of separation into excitation types, we will use a $k$-nearest neighbor classifier. The data points are first separated into a training and a test set. For each point in the test set, we consult the excitation type of its $k$ nearest neighbors in the training set. The data point is then labelled with the excitation type that the most neighbors belong to.
Here, the $k=13$ nearest neighbors are consulted, which corresponds approximately to the square root of the training set size. Contrary to other learning algorithms, there is no iterative training procedure, since the training set does not change.

\section{Results and Discussion}
\label{sec:results}

\subsection{The dataset}

The calculated excitation energies, $\Lambda$ and $\Theta$ values are available in the supplementary  information.
The energies and $\Lambda$s agree very well with previously reported values except for the PBE excitation in the third PA oligomer, where our $\Lambda$s seem swapped, and the CAM-B3LYP $D {}^1\Delta$ and $I {}^1\Sigma^-$ excitations in CO, where our $\Lambda$s are much smaller in the Turbomole results.
In the other cases any deviations are likely due to differences between different quantum chemistry programs, different grids or different convergence thresholds for the SCF iterations.

\subsection{Optimal transport results}

The regularisation parameter $\varepsilon$ in Eq. \eqref{eq:def-ot-entropic} is determined based on the maximum value of the cost function $d_\mathrm{max}$,
\begin{equation}
    \varepsilon = 10^{-\sigma} d_\mathrm{max},
    \label{eq:def_sigma}
\end{equation}
with integer $\sigma$.
In order to determine a suitable $\sigma$, we calculated $\Theta$ for the dipeptide and N$_2$ molecules at different $\sigma$, the results of which are included in the supplementary information (Tab. S1). These molecules have drastically different grid sizes and are therefore good edge cases to test $\sigma$. 
For $\sigma=4$, $\Theta$ is captured to within 2\% of its true value, which is sufficient for the purpose of this article and was therefore used in all further calculations.

\begin{figure}[htb]
\includegraphics{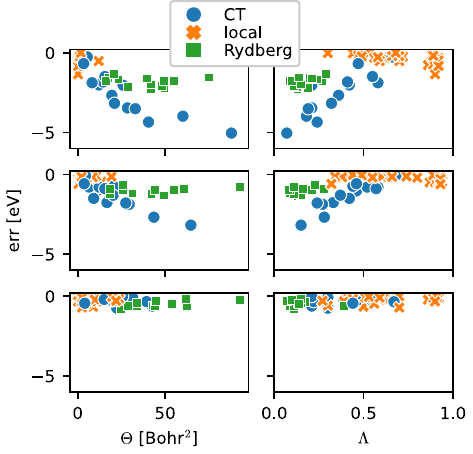}
\caption{Comparison of $\Theta$ (left) and $\Lambda$ (right) with the error in excitation energy. From top to bottom, the TDDFT functionals are PBE, B3LYP and CAM-B3LYP.}
\label{fig:diagnostic-vs-errror}
\end{figure}

\begin{figure}[htb]
    \centering
    \begin{subfigure}{.33\columnwidth}
    \centering
        \includegraphics{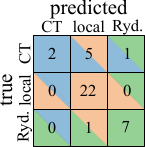}
        \subcaption{}
        \label{subfig:confusionmatrix-lambda}
    \end{subfigure}%
    \begin{subfigure}{.33\columnwidth}
    \centering
        \includegraphics{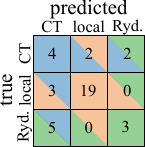}
        \subcaption{}
        \label{subfig:confusionmatrix-theta}
    \end{subfigure}%
    \begin{subfigure}{.33\columnwidth}
    \centering
        \includegraphics{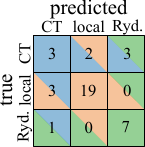}
        \subcaption{}
        \label{subfig:confusionmatrix-both}
    \end{subfigure}
    \caption{Confusion matrix of a $k$-NN classification using $\Lambda$ only (\subref{subfig:confusionmatrix-lambda}), $\Theta$ only (\subref{subfig:confusionmatrix-theta}) and both (\subref{subfig:confusionmatrix-both}).  A cell in the $i$th row and $j$th column gives the total number of points in group $i$ that were classified as group $j$. Hence, the diagonal elements are all correctly identified data points. } 
    \label{fig:confusionmatrix}
\end{figure}

To start, we compare both $\Theta$ and $\Lambda$ with the error in the electronic excitation energy (Fig.~\ref{fig:diagnostic-vs-errror}), analogously to Fig. 2 in Ref. \citenum{Peac2008}.
The two diagnostics exhibit opposite trends: $\Theta$ increases while $\Lambda$ decreases with the excitation error.
Further, $\Theta$ is generally large for Rydberg excitations and small for local excitations.
The inverse is true for $\Lambda$, which we can reason as follows:
In Rydberg excitations, the electron is excited to a very diffuse and highly delocalised orbital.
On the one hand, it will only have very small density within the support of the initial orbital and $\Lambda$ is small.
On the other hand, in order to satisfy the marginals, the transport plan $\pi$ will have non-zero entries far away from its diagonal, resulting in a large $\Theta$.
In contrast, local excitations have high overlap between the initial and final orbials, which results in a large $\Lambda$. Simultaneously, little overall density has to be transported over a small distance and therefore $\Theta$ will be small.\\
Turning our attention to CT excitations, we note that, contrary to our expectations, the CT $\Theta$ values lie between Rydberg and local values, similarly to the $\Lambda$ values.
This makes both $\Lambda$ and $\Theta$ unlikely to be suitable classifiers, which is also apparent from the results of a $k$-nearest neighbours classifier trained using either $\Theta$ or $\Lambda$ (Fig.~\ref{fig:confusionmatrix}, first two panels):
The $\Lambda$-classifier (Fig.~\ref{subfig:confusionmatrix-lambda}) performs well for both local and Rydberg excitations, but labels almost all CT excitations wrong.
The $\Theta$-classifier (Fig.~\ref{subfig:confusionmatrix-theta}) performs better for CT excitations but worse for the other excitation types.
It should be noted that we are trying to classify imbalanced data, as our data set contains significantly more local excitations than CT or Rydberg ones, which might incur additional errors.
However,  it is clear from Fig.~\ref{fig:diagnostic-vs-errror} that even imbalanced learning would not produce a very good classifier: The CT excitations are simply too spread out to allow for this.\\

Instead, we could take advantage of the opposite trends of $\Theta$ and $\Lambda$ and train a classifier on the joint $\Theta$ and $\Lambda$ values.
Its confusion matrix is shown in Fig.~\ref{subfig:confusionmatrix-both}.
The two-dimensional classification can only partially recover the advantages of each of the one-dimensional classifiers: It is as good as the $\Lambda$ classifier for Rydberg excitations, as good as the $\Theta$ classifier for local excitations (therefore worse than if we were to use $\Lambda$) and lies between the other classifiers for CT excitations.

To understand better why, let us consider the regions of overlap of CT excitations with local and Rydberg excitations, respectively.
Rydberg and CT excitations might be difficult to distinguish with the diagnostics used here because there is a property that is completely ignored: diffusivity.
We do not have the ability to discern an excitation into a diffuse orbital from one into a translated orbital.
Additional metrics that can identify Rydberg orbitals, such as the average distance of the electron to the center of the molecule or higher order moments, might provide a better classification in combination with $\Theta$.
In the same spirit, Hirose \textit{et al.} previously used a combination of $\Lambda$ and the electron-hole distance relative to the exciton size to classify excitations.\cite{Hiro2017}

One good example of coinciding CT and local excitations are the CT excitations in the HCl molecule.
Since the excitations only take place across one hydrogen-chloride bond, the associated Sinkhorn divergences are very small ($\Theta$ on the order of 4 Bohr$^2$) and similar to typical $\Theta$ values for local excitations in the same molecule.
This is in fact not the only molecule where CT excitations are associated with surprisingly low values of $\Theta$ but rather a common situation which occurs in all molecules with CT excitation considered here.\\

\begin{figure}
    \centering
    \includegraphics[width=.4\textwidth]{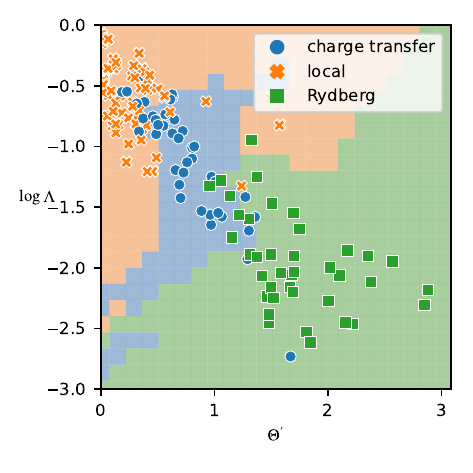}
    \caption{Combined plot of $\Theta'$ and $\log\Lambda$ for the present set of excitations. The background colours correspond to the boundaries from a $k$-NN classification.}
    \label{fig:normtheta-loglambda}
\end{figure}

We have now pointed out multiple times that there are opposite trends in $\Theta$ and $\Lambda$.
For two Gaussian densities with the same variance, there is even an explicit relation between them,
\begin{equation}
    \log\Lambda + \frac{\Theta}{4\sigma^2} = C, \label{eq:lambda-theta-relation}
\end{equation}
with a constant $C$,
which motivates us to define a new optimal transport diagnostic where the Sinkhorn divergence is normalised by the variance of the electron position,
\begin{equation}
    \Theta' = \frac{1}{\sum_{i,a} c_{ia}} \sum_{i,a} c_{ia} \frac{ S(\rho_i,\rho_a)} {\sqrt{\langle r^2\rangle_i \langle r^2 \rangle_a}}, \label{eq:normtheta}
\end{equation}
with 
\begin{equation}
    \langle r^2 \rangle_x = \int_{\mathbb{R}^3} r^2 \rho_x \mathrm{d} \boldsymbol{r},
\end{equation}
and we have assumed that the center of the molecule lies at the origin.
Note that $\Theta'$ is now dimensionless.
Plotting the new $\Theta'$ against $\log\Lambda$ (Fig.~\ref{fig:normtheta-loglambda}) reveals a close correlation between the two, which confirms that they capture the same information.
Note that Turbomole uses Gaussian basis sets, which may have an influence on the striking correlation in Fig.~\ref{fig:normtheta-loglambda} and we reserve the study of other basis functions to future work.
 
\begin{figure}[htb]
    \centering
    \begin{subfigure}{.33\columnwidth}
    \centering
        \includegraphics{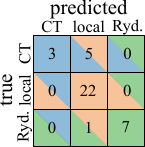}
        \subcaption{}
        \label{subfig:confusionmatrix-loglambda}
    \end{subfigure}%
    \begin{subfigure}{.33\columnwidth}
    \centering
        \includegraphics{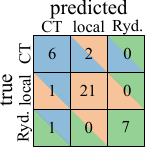}
        \subcaption{}
        \label{subfig:confusionmatrix-normtheta}
    \end{subfigure}%
    \begin{subfigure}{.33\columnwidth}
    \centering
        \includegraphics{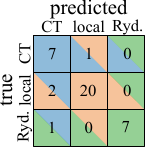}
        \subcaption{}
        \label{subfig:confusionmatrix-both-norm}
    \end{subfigure}
    \caption{Confusion matrix of a $k$-NN classification using $\log\Lambda$ only (\subref{subfig:confusionmatrix-loglambda}), $\Theta'$ only (\subref{subfig:confusionmatrix-normtheta}) and both (\subref{subfig:confusionmatrix-both-norm}). For an explanation for the confusion matrices, see caption of Fig.~\ref{fig:confusionmatrix}}.
    \label{fig:confusionmatrix-norm}
\end{figure}

Fig.~\ref{fig:confusionmatrix-norm} shows the confusion matrices for $k$-NN classifiers trained on the $\Theta'-\log\Lambda$ data set. While the $\log\Lambda$ classifier (Fig.~\ref{subfig:confusionmatrix-loglambda}) does not improve significantly on the $\Lambda$ classifier (Fig.~\ref{subfig:confusionmatrix-lambda}), we see a much better performance of the $\Theta'$ classifier (Fig.~\ref{subfig:confusionmatrix-normtheta}) on $\Theta$ (Fig.~\ref{subfig:confusionmatrix-theta}). Again, the combination does not seem to offer much additional improvement, but the current data set is too small to allow for quantitative conclusions.

\subsection{Orbital-dependence}

$\Lambda$, $\Theta$ and $\Theta'$ are calculated from the Kohn--Sham orbitals obtained in the DFT calculation.
These orbitals are not always useful to assess electronic excitations, as they can be significantly delocalised over the system.
%
%
%
Up to now, we followed the same strategy used for the $\Lambda$ diagnostic in order to have a straight-forward comparison.
However, modern developments have focused on natural transition orbitals, which give a physically more reasonable picture of the excitation on an orbital level,\cite{Mart2003} on densities and density matrices, where orbital-dependence is avoided al together,\cite{LeB2011, Baep2014, Etie2014} and on the exciton formalism.\cite{Plas2015, Baep2014, Mewe2017, Hiro2017}
The difference between orbital- and density-based descriptors has also been discussed elsewhere.\cite{Sava2017} In particular, Savarese \textit{et al.}\ conclude that natural transition orbitals are much better suited for CT diagnostics than molecular orbitals.\cite{Sava2017}
We will next briefly discuss some of these newer developments.

For the ground and excited state densities, $\rho_0$ and $\rho_X$, respectively, Le Bahers \textit{et al.}\ consider the density variation associated with a specific excitation,\cite{LeB2011}
\begin{equation}
    \Delta \rho = \rho_X - \rho_0. \label{eq:deltarho}
\end{equation}
Positive values of $\Delta \rho$ correspond to an increment in density upon excitation, negative values to a depletion.
They define a measurement for the CT length, $D_\mathrm{CT}$, as the difference between the barycenters of the positive and negative part of $\Delta \rho$.
Further, they define the spread of the regions associated with positive and negative changes in the density variation and combine them with $D_\mathrm{CT}$ to find a diagnostic for the breakdown of lower-level TDDFT functionals in CT excitations.\\

\begin{figure}
    \centering
    \includegraphics[width=.4\textwidth]{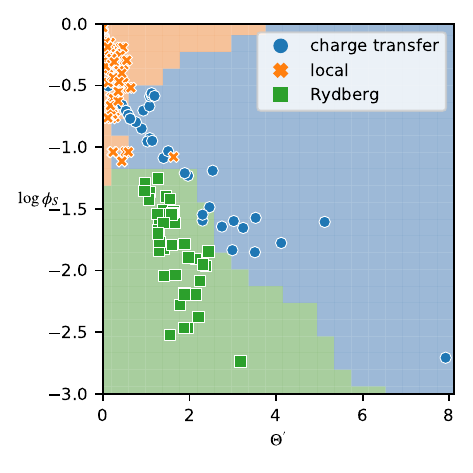}
    \caption{As Fig.~\ref{fig:normtheta-loglambda}, but for $\Theta'$ and $\phi_S$ calculated for attachment- and detachment densities. }
    \label{fig:attach-detach-classification}
\end{figure}

\begin{figure}
    \centering
    \includegraphics{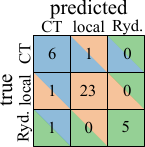}
    \caption{Confusion matrix of a $k$-NN classification using $\log\phi_\mathrm{S}$ and $\Theta'$ for attachment and detachment densities. For an explanation of the confusion matrices, see caption of Fig.~\ref{fig:confusionmatrix}}
    \label{fig:confusionmatrix-attach-detach}
\end{figure}

Alternatively, we can start from the density matrices $\hat \rho_0$ and $\hat\rho_X$ of ground and excited state. We follow Etienne \textit{et al.} in defining the difference matrix,\cite{Etie2014}
\begin{equation}
    \hat\Delta = \hat\rho_X - \hat\rho_0.
\end{equation}
If the excited state consists of singly excited Slater determinants, the difference matrix is block-diagonal with an occupied-occupied and a virtual-virtual block.\cite{Head1995}
The detachment density matrix $\hat\Gamma$ corresponds to the occupied-occupied block and the attachment density matrix $\hat\Lambda$ to the virtual-virtual block.\cite{Head1995}
From these density matrices we can define the attachment and detachment densities,
\begin{equation}
    \rho_{\tau}(\boldsymbol{r}) = \sum_{\mu, \nu} \hat\tau_{\mu \nu} \varphi_\mu(\boldsymbol{r}) \varphi_\nu^\ast(\boldsymbol{r}),
\end{equation}
where $\tau = \Gamma, \Lambda$.
Etienne \textit{et al.} define the overlap of the attachment and detachment densities,
\begin{equation}
    \phi_\mathrm{S} = \vartheta^{-1} \int_{\mathbb{R}^3} \sqrt{\rho_\Gamma(\boldsymbol{r}) \rho_\Lambda(\boldsymbol{r})} \mathrm{d} \boldsymbol{r},
\end{equation}
where $\vartheta = \frac{1}{2}\int_{\mathbb{R}^3} \sum_{\tau \in \{ \Lambda, \Gamma \}} \rho_\tau(\boldsymbol{r}) \mathrm{d}\boldsymbol{r}$,
and show that they can make similar conclusions as for the $\Lambda$ diagnostic, but with the benefit of orbital-independence.\cite{Etie2014}\\
Since the attachment and detachment densities hold the same mass, it is straight-forward to apply our optimal transport formalism.
We simply evaluate the Sinkhorn divergence for the attachment and detachment densities, for which our diagnostic can be written as
\begin{equation}
    \Theta' = \frac{S(\rho_\Lambda, \rho_\Gamma)}{\sqrt{\langle r^2 \rangle_\Lambda \langle r^2 \rangle_\Gamma}}. \label{eq:normtheta-densities}
\end{equation}
Note that, if the attachment and detachment densities have the same overall shape, the Sinkhorn divergence would reduce to the distance between the centroids of the densities (for small enough $\varepsilon$) and would be closely related to $D_\mathrm{CT}$.
In Fig.~\ref{fig:normtheta-loglambda}, we compared $\Theta'$ to the overlap-based $\Lambda$. In a similar spirit, we compare $\Theta'$ to $\phi_S$ in the context of densities (see Fig.~\ref{fig:attach-detach-classification}).
Compared to Fig.~\ref{fig:normtheta-loglambda}, the CT and Rydberg excitations are clustered very differently, showing the orbital dependence of our previous results. At the same overlap value, the Sinkhorn divergence is larger for CT excitations than for Rydberg excitations, leading to a diagonal separating plane, whereas in Fig.~\ref{fig:normtheta-loglambda} the CT excitations were squeezed between the other two regions. The improved separation between the excitation types is also reflected in the associated confusion matrix (Fig. \ref{fig:confusionmatrix-attach-detach}), where we reach a near perfect classification.

In future work, $\Theta'$ should therefore be applied either in the manner of Eq.~\eqref{eq:normtheta-densities} or in the manner of Eq.~\eqref{eq:normtheta}, but with natural transition orbitals.
Both of these strategies mitigate the extent of orbital dependence.

In Fig.~\ref{fig:attach-detach-classification}, there are still data points that lie close to the boundary between excitation types.
In particular in these boundary cases, it is sensible to think of these diagnostics as measuring the extent of an excitation character rather than assigning a definite label.
In such cases it may be useful to introduce a fractional decomposition into excitation characters.
We can now also question the labels in our data set based on the results in Fig. \ref{fig:attach-detach-classification}. For example, the ${}^1B_1$ excitation in the formaldehyde ($\Theta' = 1.6$, $\log\phi_S = -1.0$) is labelled as a local excitation in our data set, but clearly lies the CT labelled region, an indication that the excitation carries significant CT character.

Finally, let us touch on diagnostics derived from the exciton formalism. The electron and hole sizes from the exciton formalism are useful to analyse Rydberg states.\cite{Plas2015} Further, the exciton size $d_\mathrm{exc}$ has been shown to be large for CT and Rydberg excitations.\cite{Baep2014}
$d_\mathrm{exc}$ also works well for centrosymmetric molecules,\cite{Mewe2017} which have proven challenging for other diagnostics like the electron-hole distance $d_\mathrm{eh}$ and the CT length $D_\mathrm{CT}$.\cite{LeB2011,Hiro2017}
It can be expected that the optimal transport diagnostics would also struggle with centrosymmetric molecules.

\section{Conclusion}
\label{sec:conclusion}

There is currently no reliable way to classify electronic excitations into CT, Rydberg and local excitations.
In this report we studied two diagnostics, the overlap-based $\Lambda$,\cite{Peac2008} the optimal transport-based $\Theta$ and their combination to classify an electronic excitation via its $k$-nearest neighbours.
There are strengths and issues in both $\Lambda$ and $\Theta$ and their combination gives a compromise, but it still does not give a viable classifier.
We argued that CT excitations can  be similar to local excitations in $\Theta$ and $\Lambda$ if the donor and acceptor are close together, especially in small molecules.
We further argued that $\Theta$ and $\Lambda$ cannot distinguish between CT and Rydberg excitations because neither of them captures the diffusivity of the final orbital in a Rydberg excitation.

We then showed that there is a relation between $\log \Lambda$ and a modified diagnostic $\Theta'$, which is a length scale-corrected version of $\Theta$. The combination of $\Theta'$ and $\log\Lambda$ gives the best classifier explored in this work which is purely based on KS orbitals, but more studies are necessary into whether $\Theta'$ is also a sensible diagnostic in programs that use non-Gaussian orbitals.
We also discuss the orbital-dependence in $\Theta'$ and propose a density matrix-based version, which compares attachment and detachment densities. We compare this modification to the overlap between the two densities,\cite{Etie2014} and observe a very different clustering of the data points which leads to a significant improvement in the classification. This suggests that the orbital choice has a significant impact.
In future work, we also want to compare $\Theta'$ diagnostics with other modern CT diagnostics,\cite{Baep2014, Plas2015, LeB2011, Hiro2017,Mewe2017, Mewe2015, Mewe2019, Guid2013, Dev2012, Moor2015} to learn more about its strengths and weaknesses. Modern diagnostics will additionally help in the differentiation between CT and Rydberg excitations.

While this work was being finalised, a study that shares similar ideas has appeared.\cite{Wang2023}
The study focuses on the density difference defined in Eq. \eqref{eq:deltarho} and looks at the optimal transport between the regions of density depletion ($\Delta \rho < 0$) and density enhancement ($\Delta \rho > 0$).
We reach similar conclusions, but we have also key differences, like the use of the entropic regularisation to accelerate calculations and a different look at possible classifications.

\begin{acknowledgement}

The authors thank Stijn de Baerdemacker for pointing out Eq.~\eqref{eq:lambda-theta-relation}, David Tew for helpful comments and advice on the Turbomole calculations, Joe Cooper for helpful discussions about Rydberg excitations, Giulia Luise for helpful discussions about the Sinkhorn divergence as well the \texttt{geomloss} package, Derk Kooi for advice on the initial version of the Sinkhorn algorithm, and Hans Chan for helpful suggestions on the manuscript. Annina Lieberherr was supported by a Berrow Foundation Lord Florey Scholarship and, during the initial phase of this work, by the University Research Fellowship at the Vrije Universiteit Amsterdam.

\end{acknowledgement}

\begin{suppinfo}

The supplementary information shows the molecules considered here, the convergence of $\Theta$ with the regularisation parameter $\sigma$ and a link to a github repository. The data needed to reproduce the results here is contained in the file \texttt{si\_data.xlsx}.

\end{suppinfo}

\bibliography{ot}

\providecommand{\latin}[1]{#1}
\makeatletter
\providecommand{\doi}
  {\begingroup\let\do\@makeother\dospecials
  \catcode`\{=1 \catcode`\}=2 \doi@aux}
\providecommand{\doi@aux}[1]{\endgroup\texttt{#1}}
\makeatother
\providecommand*\mcitethebibliography{\thebibliography}
\csname @ifundefined\endcsname{endmcitethebibliography}  {\let\endmcitethebibliography\endthebibliography}{}
\begin{mcitethebibliography}{39}
\providecommand*\natexlab[1]{#1}
\providecommand*\mciteSetBstSublistMode[1]{}
\providecommand*\mciteSetBstMaxWidthForm[2]{}
\providecommand*\mciteBstWouldAddEndPuncttrue
  {\def\EndOfBibitem{\unskip.}}
\providecommand*\mciteBstWouldAddEndPunctfalse
  {\let\EndOfBibitem\relax}
\providecommand*\mciteSetBstMidEndSepPunct[3]{}
\providecommand*\mciteSetBstSublistLabelBeginEnd[3]{}
\providecommand*\EndOfBibitem{}
\mciteSetBstSublistMode{f}
\mciteSetBstMaxWidthForm{subitem}{(\alph{mcitesubitemcount})}
\mciteSetBstSublistLabelBeginEnd
  {\mcitemaxwidthsubitemform\space}
  {\relax}
  {\relax}

\bibitem[Scholes(2017)]{Scho2017}
Scholes,~G.~D. Introduction: Light harvesting. \emph{Chem. Rev.} \textbf{2017}, \emph{117}, 247--248\relax
\mciteBstWouldAddEndPuncttrue
\mciteSetBstMidEndSepPunct{\mcitedefaultmidpunct}
{\mcitedefaultendpunct}{\mcitedefaultseppunct}\relax
\EndOfBibitem
\bibitem[Hagfeldt \latin{et~al.}(2010)Hagfeldt, Boschloo, Sun, Kloo, and Pettersson]{Hagf2010}
Hagfeldt,~A.; Boschloo,~G.; Sun,~L.; Kloo,~L.; Pettersson,~H. Dye-sensitized solar cells. \emph{Chem. Rev.} \textbf{2010}, \emph{110}, 6595--6663\relax
\mciteBstWouldAddEndPuncttrue
\mciteSetBstMidEndSepPunct{\mcitedefaultmidpunct}
{\mcitedefaultendpunct}{\mcitedefaultseppunct}\relax
\EndOfBibitem
\bibitem[Canc{\`{e}}s and Friesecke(2023)Canc{\`{e}}s, and Friesecke]{Canc2023}
Canc{\`{e}}s,~E., Friesecke,~G., Eds. \emph{Density functional theory}; Springer International Publishing, 2023\relax
\mciteBstWouldAddEndPuncttrue
\mciteSetBstMidEndSepPunct{\mcitedefaultmidpunct}
{\mcitedefaultendpunct}{\mcitedefaultseppunct}\relax
\EndOfBibitem
\bibitem[Medvedev \latin{et~al.}(2017)Medvedev, Bushmarinov, Sun, Perdew, and Lyssenko]{Medv2017}
Medvedev,~M.~G.; Bushmarinov,~I.~S.; Sun,~J.; Perdew,~J.~P.; Lyssenko,~K.~A. Density functional theory is straying from the path toward the exact functional. \emph{Science} \textbf{2017}, \emph{355}, 49--52\relax
\mciteBstWouldAddEndPuncttrue
\mciteSetBstMidEndSepPunct{\mcitedefaultmidpunct}
{\mcitedefaultendpunct}{\mcitedefaultseppunct}\relax
\EndOfBibitem
\bibitem[Ullrich(2012)]{Ullr2006}
Ullrich,~C.~A. \emph{Time dependent density functional theory}; Oxford Graduate Texts, 2012\relax
\mciteBstWouldAddEndPuncttrue
\mciteSetBstMidEndSepPunct{\mcitedefaultmidpunct}
{\mcitedefaultendpunct}{\mcitedefaultseppunct}\relax
\EndOfBibitem
\bibitem[Maitra(2017)]{Mait2017}
Maitra,~N.~T. Charge transfer in time-dependent density functional theory. \emph{J. Phys.: Condens. Matter} \textbf{2017}, \emph{29}, 423001\relax
\mciteBstWouldAddEndPuncttrue
\mciteSetBstMidEndSepPunct{\mcitedefaultmidpunct}
{\mcitedefaultendpunct}{\mcitedefaultseppunct}\relax
\EndOfBibitem
\bibitem[Peach \latin{et~al.}(2008)Peach, Benfield, Helgaker, and Tozer]{Peac2008}
Peach,~M. J.~G.; Benfield,~P.; Helgaker,~T.; Tozer,~D.~J. Excitation energies in density functional theory: An evaluation and a diagnostic test. \emph{J. Chem. Phys.} \textbf{2008}, \emph{128}, 044118\relax
\mciteBstWouldAddEndPuncttrue
\mciteSetBstMidEndSepPunct{\mcitedefaultmidpunct}
{\mcitedefaultendpunct}{\mcitedefaultseppunct}\relax
\EndOfBibitem
\bibitem[Mewes \latin{et~al.}(2015)Mewes, Plasser, and Dreuw]{Mewe2015}
Mewes,~S.~A.; Plasser,~F.; Dreuw,~A. Communication: Exciton analysis in time-dependent density functional theory: How functionals shape excited-state characters. \emph{J. Chem. Phys.} \textbf{2015}, \emph{143}, 171101\relax
\mciteBstWouldAddEndPuncttrue
\mciteSetBstMidEndSepPunct{\mcitedefaultmidpunct}
{\mcitedefaultendpunct}{\mcitedefaultseppunct}\relax
\EndOfBibitem
\bibitem[Hirose \latin{et~al.}(2017)Hirose, Noguchi, and Sugino]{Hiro2017}
Hirose,~D.; Noguchi,~Y.; Sugino,~O. Quantitative characterization of exciton from \emph{GW}+{B}ethe-{S}alpeter calculation. \emph{J. Chem. Phys.} \textbf{2017}, \emph{146}, 044303\relax
\mciteBstWouldAddEndPuncttrue
\mciteSetBstMidEndSepPunct{\mcitedefaultmidpunct}
{\mcitedefaultendpunct}{\mcitedefaultseppunct}\relax
\EndOfBibitem
\bibitem[Dev \latin{et~al.}(2012)Dev, Agrawal, and English]{Dev2012}
Dev,~P.; Agrawal,~S.; English,~N.~J. Determining the appropriate exchange-correlation functional for time-dependent density functional theory studies of charge-transfer excitations in organic dyes. \emph{J. Chem. Phys.} \textbf{2012}, \emph{136}, 224301\relax
\mciteBstWouldAddEndPuncttrue
\mciteSetBstMidEndSepPunct{\mcitedefaultmidpunct}
{\mcitedefaultendpunct}{\mcitedefaultseppunct}\relax
\EndOfBibitem
\bibitem[Leang \latin{et~al.}(2012)Leang, Zahariev, and Gordon]{Lean2012}
Leang,~S.~S.; Zahariev,~F.; Gordon,~M.~S. Benchmarking the performance of time-dependent density functional methods. \emph{J. Chem. Phys.} \textbf{2012}, \emph{136}, 104101\relax
\mciteBstWouldAddEndPuncttrue
\mciteSetBstMidEndSepPunct{\mcitedefaultmidpunct}
{\mcitedefaultendpunct}{\mcitedefaultseppunct}\relax
\EndOfBibitem
\bibitem[Kornobis \latin{et~al.}(2011)Kornobis, Kumar, Wong, Lodowski, Jaworska, Andruni{\'{o}}w, Ruud, and Kozlowski]{Korn2011}
Kornobis,~K.; Kumar,~N.; Wong,~B.~M.; Lodowski,~P.; Jaworska,~M.; Andruni{\'{o}}w,~T.; Ruud,~K.; Kozlowski,~P.~M. Electronically excited states of vitamin {B}12: Benchmark calculations including time-dependent density functional theory and correlated ab initio methods. \emph{J. Phys. Chem. A} \textbf{2011}, \emph{115}, 1280--1292\relax
\mciteBstWouldAddEndPuncttrue
\mciteSetBstMidEndSepPunct{\mcitedefaultmidpunct}
{\mcitedefaultendpunct}{\mcitedefaultseppunct}\relax
\EndOfBibitem
\bibitem[Guido \latin{et~al.}(2013)Guido, Cortona, Mennucci, and Adamo]{Guid2013}
Guido,~C.~A.; Cortona,~P.; Mennucci,~B.; Adamo,~C. On the metric of charge transfer molecular excitations: A simple chemical descriptor. \emph{J. Chem. Theory Comput.} \textbf{2013}, \emph{9}, 3118--3126\relax
\mciteBstWouldAddEndPuncttrue
\mciteSetBstMidEndSepPunct{\mcitedefaultmidpunct}
{\mcitedefaultendpunct}{\mcitedefaultseppunct}\relax
\EndOfBibitem
\bibitem[Mewes and Dreuw(2019)Mewes, and Dreuw]{Mewe2019}
Mewes,~S.~A.; Dreuw,~A. Density-based descriptors and exciton analyses for visualizing and understanding the electronic structure of excited states. \emph{Phys. Chem. Chem. Phys.} \textbf{2019}, \emph{21}, 2843--2856\relax
\mciteBstWouldAddEndPuncttrue
\mciteSetBstMidEndSepPunct{\mcitedefaultmidpunct}
{\mcitedefaultendpunct}{\mcitedefaultseppunct}\relax
\EndOfBibitem
\bibitem[Moore \latin{et~al.}(2015)Moore, Sun, Govind, Kowalski, and Autschbach]{Moor2015}
Moore,~B.; Sun,~H.; Govind,~N.; Kowalski,~K.; Autschbach,~J. Charge-transfer versus charge-transfer-like excitations revisited. \emph{J. Chem. Theory Comput.} \textbf{2015}, \emph{11}, 3305--3320\relax
\mciteBstWouldAddEndPuncttrue
\mciteSetBstMidEndSepPunct{\mcitedefaultmidpunct}
{\mcitedefaultendpunct}{\mcitedefaultseppunct}\relax
\EndOfBibitem
\bibitem[Kohn and Sham(1965)Kohn, and Sham]{Kohn1965}
Kohn,~W.; Sham,~L.~J. Self-consistent equations including exchange and correlation effects. \emph{Phys. Rev.} \textbf{1965}, \emph{140}, A1133--A1138\relax
\mciteBstWouldAddEndPuncttrue
\mciteSetBstMidEndSepPunct{\mcitedefaultmidpunct}
{\mcitedefaultendpunct}{\mcitedefaultseppunct}\relax
\EndOfBibitem
\bibitem[Casida(1995)]{Casi1995}
Casida,~M.~E. In \emph{Recent Advances in Density Functional Methods}; Chong,~D.~P., Ed.; World Scientific, 1995; Chapter Time-dependent density functional response theory for molecules, pp 155--192\relax
\mciteBstWouldAddEndPuncttrue
\mciteSetBstMidEndSepPunct{\mcitedefaultmidpunct}
{\mcitedefaultendpunct}{\mcitedefaultseppunct}\relax
\EndOfBibitem
\bibitem[Toulouse(2022)]{Toul2022}
Toulouse,~J. In \emph{Density Functional Theory}; Canc{\`{e}}s,~E., Friesecke,~G., Eds.; Springer International Publishing, 2022; Chapter Review of approximations for the exchange-correlation energy in density-functional theory, pp 1--90\relax
\mciteBstWouldAddEndPuncttrue
\mciteSetBstMidEndSepPunct{\mcitedefaultmidpunct}
{\mcitedefaultendpunct}{\mcitedefaultseppunct}\relax
\EndOfBibitem
\bibitem[Peyr{\'{e}} and Cuturi(2019)Peyr{\'{e}}, and Cuturi]{Peyr2019}
Peyr{\'{e}},~G.; Cuturi,~M. Computational optimal transport: With applications to data science. \emph{Foundations and Trends{\textregistered} in Machine Learning} \textbf{2019}, \emph{11}, 355--607\relax
\mciteBstWouldAddEndPuncttrue
\mciteSetBstMidEndSepPunct{\mcitedefaultmidpunct}
{\mcitedefaultendpunct}{\mcitedefaultseppunct}\relax
\EndOfBibitem
\bibitem[Cuturi(2013)]{Cutu2013}
Cuturi,~M. Sinkhorn distances: Lightspeed computation of Optimal Transport. Advances in Neural Information Processing Systems. 2013\relax
\mciteBstWouldAddEndPuncttrue
\mciteSetBstMidEndSepPunct{\mcitedefaultmidpunct}
{\mcitedefaultendpunct}{\mcitedefaultseppunct}\relax
\EndOfBibitem
\bibitem[Ramdas \latin{et~al.}(2017)Ramdas, Trillos, and Cuturi]{Ramd2017}
Ramdas,~A.; Trillos,~N.; Cuturi,~M. On Wasserstein two-sample testing and related families of nonparametric tests. \emph{Entropy} \textbf{2017}, \emph{19}, 47\relax
\mciteBstWouldAddEndPuncttrue
\mciteSetBstMidEndSepPunct{\mcitedefaultmidpunct}
{\mcitedefaultendpunct}{\mcitedefaultseppunct}\relax
\EndOfBibitem
\bibitem[Balasubramani \latin{et~al.}(2020)Balasubramani, Chen, Coriani, Diedenhofen, Frank, Franzke, Furche, Grotjahn, Harding, Hättig, Hellweg, Helmich-Paris, Holzer, Huniar, Kaupp, Khah, Khani, Müller, Mack, Nguyen, Parker, Perlt, Rappoport, Reiter, Roy, Rückert, Schmitz, Sierka, Tapavicza, Tew, van Wüllen, Voora, Weigend, Wody{\'{n}}ski, and Yu]{Bala2020}
Balasubramani,~S.~G.; Chen,~G.~P.; Coriani,~S.; Diedenhofen,~M.; Frank,~M.~S.; Franzke,~Y.~J.; Furche,~F.; Grotjahn,~R.; Harding,~M.~E.; Hättig,~C. \latin{et~al.}  {TURBOMOLE}: Modular program suite for \emph{ab initio} quantum-chemical and condensed-matter simulations. \emph{J. Chem. Phys.} \textbf{2020}, \emph{152}, 184107\relax
\mciteBstWouldAddEndPuncttrue
\mciteSetBstMidEndSepPunct{\mcitedefaultmidpunct}
{\mcitedefaultendpunct}{\mcitedefaultseppunct}\relax
\EndOfBibitem
\bibitem[Dunning(1989)]{Dunn1989}
Dunning,~T.~H. Gaussian basis sets for use in correlated molecular calculations. I. The atoms boron through neon and hydrogen. \emph{J. Chem. Phys.} \textbf{1989}, \emph{90}, 1007--1023\relax
\mciteBstWouldAddEndPuncttrue
\mciteSetBstMidEndSepPunct{\mcitedefaultmidpunct}
{\mcitedefaultendpunct}{\mcitedefaultseppunct}\relax
\EndOfBibitem
\bibitem[Kendall \latin{et~al.}(1992)Kendall, Dunning, and Harrison]{Kend1992}
Kendall,~R.~A.; Dunning,~T.~H.; Harrison,~R.~J. Electron affinities of the first-row atoms revisited. Systematic basis sets and wave functions. \emph{J. Chem. Phys.} \textbf{1992}, \emph{96}, 6796--6806\relax
\mciteBstWouldAddEndPuncttrue
\mciteSetBstMidEndSepPunct{\mcitedefaultmidpunct}
{\mcitedefaultendpunct}{\mcitedefaultseppunct}\relax
\EndOfBibitem
\bibitem[Woon and Dunning(1994)Woon, and Dunning]{Woon1994}
Woon,~D.~E.; Dunning,~T.~H. Gaussian basis sets for use in correlated molecular calculations. {IV}. Calculation of static electrical response properties. \emph{J. Chem. Phys.} \textbf{1994}, \emph{100}, 2975--2988\relax
\mciteBstWouldAddEndPuncttrue
\mciteSetBstMidEndSepPunct{\mcitedefaultmidpunct}
{\mcitedefaultendpunct}{\mcitedefaultseppunct}\relax
\EndOfBibitem
\bibitem[Schuchardt \latin{et~al.}(2007)Schuchardt, Didier, Elsethagen, Sun, Gurumoorthi, Chase, Li, and Windus]{Schu2007}
Schuchardt,~K.~L.; Didier,~B.~T.; Elsethagen,~T.; Sun,~L.; Gurumoorthi,~V.; Chase,~J.; Li,~J.; Windus,~T.~L. Basis set exchange: A community database for computational sciences. \emph{J. Chem. Inf. Model.} \textbf{2007}, \emph{47}, 1045–1052\relax
\mciteBstWouldAddEndPuncttrue
\mciteSetBstMidEndSepPunct{\mcitedefaultmidpunct}
{\mcitedefaultendpunct}{\mcitedefaultseppunct}\relax
\EndOfBibitem
\bibitem[Pritchard \latin{et~al.}(2019)Pritchard, Altarawy, Didier, Gibsom, and Windus]{Prit2019}
Pritchard,~B.~P.; Altarawy,~D.; Didier,~B.; Gibsom,~T.~D.; Windus,~T.~L. New basis set exchange: An open, up-to-date resource for the molecular sciences community. \emph{J. Chem. Inf. Model} \textbf{2019}, \emph{59}, 4814--4820\relax
\mciteBstWouldAddEndPuncttrue
\mciteSetBstMidEndSepPunct{\mcitedefaultmidpunct}
{\mcitedefaultendpunct}{\mcitedefaultseppunct}\relax
\EndOfBibitem
\bibitem[Eichkorn \latin{et~al.}(1997)Eichkorn, Weigend, Treutler, and Ahlrichs]{Eich1997}
Eichkorn,~K.; Weigend,~F.; Treutler,~O.; Ahlrichs,~R. Auxiliary basis sets for main row atoms and transition metals and their use to approximate Coulomb potentials. \emph{Theor. Chem. Acc.} \textbf{1997}, \emph{97}, 119--124\relax
\mciteBstWouldAddEndPuncttrue
\mciteSetBstMidEndSepPunct{\mcitedefaultmidpunct}
{\mcitedefaultendpunct}{\mcitedefaultseppunct}\relax
\EndOfBibitem
\bibitem[Feydy \latin{et~al.}(2019)Feydy, S\'{e}journ\'{e}, Vialard, Amari, Trouve, and Peyr\'{e}]{Feyd2019}
Feydy,~J.; S\'{e}journ\'{e},~T.; Vialard,~F.-X.; Amari,~S.-i.; Trouve,~A.; Peyr\'{e},~G. Interpolating between optimal transport and MMD using Sinkhorn divergences. Proceedings of the Twenty-Second International Conference on Artificial Intelligence and Statistics. 2019; pp 2681--2690\relax
\mciteBstWouldAddEndPuncttrue
\mciteSetBstMidEndSepPunct{\mcitedefaultmidpunct}
{\mcitedefaultendpunct}{\mcitedefaultseppunct}\relax
\EndOfBibitem
\bibitem[Martin(2003)]{Mart2003}
Martin,~R.~L. Natural transition orbitals. \emph{J. Chem. Phys.} \textbf{2003}, \emph{118}, 4775--4777\relax
\mciteBstWouldAddEndPuncttrue
\mciteSetBstMidEndSepPunct{\mcitedefaultmidpunct}
{\mcitedefaultendpunct}{\mcitedefaultseppunct}\relax
\EndOfBibitem
\bibitem[Le~Bahers \latin{et~al.}(2011)Le~Bahers, Adamo, and Ciofini]{LeB2011}
Le~Bahers,~T.; Adamo,~C.; Ciofini,~I. A qualitative index of spatial extent in charge-transfer excitations. \emph{J. Chem. Theory Comput.} \textbf{2011}, \emph{7}, 2498--2506\relax
\mciteBstWouldAddEndPuncttrue
\mciteSetBstMidEndSepPunct{\mcitedefaultmidpunct}
{\mcitedefaultendpunct}{\mcitedefaultseppunct}\relax
\EndOfBibitem
\bibitem[Bäppler \latin{et~al.}(2014)Bäppler, Plasser, Wormit, and Dreuw]{Baep2014}
Bäppler,~S.~A.; Plasser,~F.; Wormit,~M.; Dreuw,~A. Exciton analysis of many-body wave functions: Bridging the gap between the quasiparticle and molecular orbital pictures. \emph{Phys. Rev. A} \textbf{2014}, \emph{90}, 052521\relax
\mciteBstWouldAddEndPuncttrue
\mciteSetBstMidEndSepPunct{\mcitedefaultmidpunct}
{\mcitedefaultendpunct}{\mcitedefaultseppunct}\relax
\EndOfBibitem
\bibitem[Etienne \latin{et~al.}(2014)Etienne, Assfeld, and Monari]{Etie2014}
Etienne,~T.; Assfeld,~X.; Monari,~A. Toward a quantitative assessment of electronic transitions’ charge-transfer character. \emph{J. Chem. Theory Comput.} \textbf{2014}, \emph{10}, 3896--3905\relax
\mciteBstWouldAddEndPuncttrue
\mciteSetBstMidEndSepPunct{\mcitedefaultmidpunct}
{\mcitedefaultendpunct}{\mcitedefaultseppunct}\relax
\EndOfBibitem
\bibitem[Plasser \latin{et~al.}(2015)Plasser, Thomitzni, Bäppler, Wenzel, Rehn, Wormit, and Dreuw]{Plas2015}
Plasser,~F.; Thomitzni,~B.; Bäppler,~S.~A.; Wenzel,~J.; Rehn,~D.~R.; Wormit,~M.; Dreuw,~A. Statistical analysis of electronic excitation processes: Spatial location, compactness, charge transfer, and electron-hole correlation. \emph{J. Comput. Chem.} \textbf{2015}, \emph{36}, 1609--1620\relax
\mciteBstWouldAddEndPuncttrue
\mciteSetBstMidEndSepPunct{\mcitedefaultmidpunct}
{\mcitedefaultendpunct}{\mcitedefaultseppunct}\relax
\EndOfBibitem
\bibitem[Mewes \latin{et~al.}(2017)Mewes, Plasser, and Dreuw]{Mewe2017}
Mewes,~S.~A.; Plasser,~F.; Dreuw,~A. Universal exciton size in organic polymers is determined by nonlocal orbital exchange in time-dependent density functional theory. \emph{J. Phys. Chem. Lett.} \textbf{2017}, \emph{8}, 1205--1210\relax
\mciteBstWouldAddEndPuncttrue
\mciteSetBstMidEndSepPunct{\mcitedefaultmidpunct}
{\mcitedefaultendpunct}{\mcitedefaultseppunct}\relax
\EndOfBibitem
\bibitem[Savarese \latin{et~al.}(2017)Savarese, Guido, Brémond, Ciofini, and Adamo]{Sava2017}
Savarese,~M.; Guido,~C.~A.; Brémond,~E.; Ciofini,~I.; Adamo,~C. Metrics for molecular electronic excitations: A comparison between orbital- and density-based descriptors. \emph{J. Phys. Chem. A} \textbf{2017}, \emph{121}, 7543--7549\relax
\mciteBstWouldAddEndPuncttrue
\mciteSetBstMidEndSepPunct{\mcitedefaultmidpunct}
{\mcitedefaultendpunct}{\mcitedefaultseppunct}\relax
\EndOfBibitem
\bibitem[Head-Gordon \latin{et~al.}(1995)Head-Gordon, Grana, Maurice, and White]{Head1995}
Head-Gordon,~M.; Grana,~A.~M.; Maurice,~D.; White,~C.~A. Analysis of electronic transitions as the difference of electron attachment and detachment densities. \emph{J. Phys. Chem.} \textbf{1995}, \emph{99}, 14261--14270\relax
\mciteBstWouldAddEndPuncttrue
\mciteSetBstMidEndSepPunct{\mcitedefaultmidpunct}
{\mcitedefaultendpunct}{\mcitedefaultseppunct}\relax
\EndOfBibitem
\bibitem[Wang \latin{et~al.}(2023)Wang, Liang, and Head-Gordon]{Wang2023}
Wang,~Z.; Liang,~J.; Head-Gordon,~M. Earth mover's distance as a metric to evaluate the extent of charge transfer in excitations using discretized real-space densities. \emph{J. Chem. Theory Comput.} \textbf{2023}, \emph{19}, 7704--7714\relax
\mciteBstWouldAddEndPuncttrue
\mciteSetBstMidEndSepPunct{\mcitedefaultmidpunct}
{\mcitedefaultendpunct}{\mcitedefaultseppunct}\relax
\EndOfBibitem
\end{mcitethebibliography}


\providecommand{\latin}[1]{#1}
\makeatletter
\providecommand{\doi}
  {\begingroup\let\do\@makeother\dospecials
  \catcode`\{=1 \catcode`\}=2 \doi@aux}
\providecommand{\doi@aux}[1]{\endgroup\texttt{#1}}
\makeatother
\providecommand*\mcitethebibliography{\thebibliography}
\csname @ifundefined\endcsname{endmcitethebibliography}  {\let\endmcitethebibliography\endthebibliography}{}
\begin{mcitethebibliography}{2}
\providecommand*\natexlab[1]{#1}
\providecommand*\mciteSetBstSublistMode[1]{}
\providecommand*\mciteSetBstMaxWidthForm[2]{}
\providecommand*\mciteBstWouldAddEndPuncttrue
  {\def\EndOfBibitem{\unskip.}}
\providecommand*\mciteBstWouldAddEndPunctfalse
  {\let\EndOfBibitem\relax}
\providecommand*\mciteSetBstMidEndSepPunct[3]{}
\providecommand*\mciteSetBstSublistLabelBeginEnd[3]{}
\providecommand*\EndOfBibitem{}
\mciteSetBstSublistMode{f}
\mciteSetBstMaxWidthForm{subitem}{(\alph{mcitesubitemcount})}
\mciteSetBstSublistLabelBeginEnd
  {\mcitemaxwidthsubitemform\space}
  {\relax}
  {\relax}

\bibitem[Peach \latin{et~al.}(2008)Peach, Benfield, Helgaker, and Tozer]{Peac2008}
Peach,~M. J.~G.; Benfield,~P.; Helgaker,~T.; Tozer,~D.~J. \emph{J. Chem. Phys.} \textbf{2008}, \emph{128}, 044118\relax
\mciteBstWouldAddEndPuncttrue
\mciteSetBstMidEndSepPunct{\mcitedefaultmidpunct}
{\mcitedefaultendpunct}{\mcitedefaultseppunct}\relax
\EndOfBibitem
\end{mcitethebibliography}

\begin{tocentry}
    \includegraphics{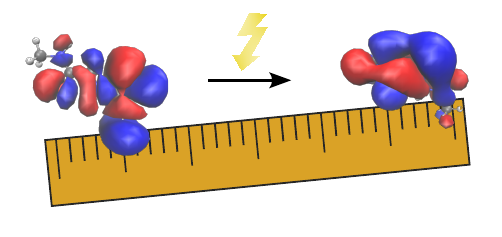}
\end{tocentry}

\end{document}


\section*{Supporting Information}

\subsection*{Molecules}

\begin{figure}[htb]
    \label{fig:molecules}
    \centering
    \begin{subfigure}{.3\textwidth}
    \centering
    \includegraphics[width=3.5cm]{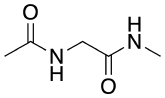}
    \subcaption*{Dipeptide}
    \end{subfigure}%
    \begin{subfigure}{.3\textwidth}
    \centering
    \includegraphics[width=4cm]{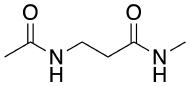}
    \subcaption*{$\beta$-dipeptide}
    \end{subfigure}%
    \begin{subfigure}{.3\textwidth}
    \centering
    \includegraphics[width=4.7cm]{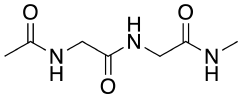}
    \subcaption*{Tripeptide}
    \end{subfigure}

    \begin{subfigure}[t]{.3\textwidth}
    \centering
        \includegraphics[width=3cm]{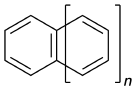}
        \subcaption*{Acenes ($n=1$-5)}
    \end{subfigure}%
    \begin{subfigure}[t]{.3\textwidth}
    \centering
        \includegraphics[width=3cm]{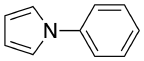}
        \subcaption*{N-phenylpyrrole (PP)}
    \end{subfigure}%
    \begin{subfigure}[t]{.4\textwidth}
    \centering
        \includegraphics[width=3.7cm]{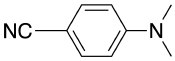}
        \subcaption*{4-(N,N-dimethylamino)benzonitrile (DMABN)}
    \end{subfigure}

    \begin{subfigure}[t]{.35\textwidth}
    \centering
        \includegraphics[width=2.2cm]{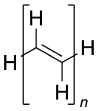}
        \subcaption*{Polyacetylene oligomers ($n=2$-5)}
    \end{subfigure}
    
    \begin{subfigure}[b]{.1\textwidth}
    \centering
        \includegraphics[width=.7cm]{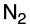}
    \end{subfigure}
    \begin{subfigure}[b]{.1\textwidth}
    \centering
        \includegraphics[width=.9cm]{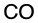}
    \end{subfigure}
    \begin{subfigure}[b]{.2\textwidth}
    \centering
        \includegraphics[width=1.7cm]{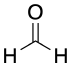}
    \end{subfigure}
    \begin{subfigure}[b]{.1\textwidth}
    \centering
        \includegraphics[width=1.cm]{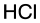}
    \end{subfigure}
    
    \caption{Molecules studied by Peach \textit{et al.} and here.\cite{Peac2008}}
\end{figure}
\clearpage

\subsection*{Convergence with respect to entropic regularisation}

\begin{table}[htb]
\label{tab:thetaconv_reg}

\caption{Convergence of $\Theta$ with regularisation parameter $\varepsilon = 10^{-\sigma} d_\mathrm{max}$, where $d_\mathrm{max}$ is the maximum possible distance for a molecular grid. The dipeptide shows the trend for the cc-pVTZ basis set, the N$_2$ molecule for the $d$-aug-cc-pVTZ basis set. }
\centering
    \begin{tabular}{ll|rrrr}
        Excitation & Functional & \multicolumn{4}{c}{$\sigma$} \\
        & & 3 & 4 & 5 & 6 \\
        \hline
        dipeptide,  $n_1\rightarrow\pi_2^\ast$ & PBE & 2\textbf{7.93} & 28.4\textbf{7} & 28.48 & \np{28.48390886} \\
        & B3LYP & 16.\textbf{22} & 16.7\textbf{7} & \np{16.784085} & \np{16.784177} \\
        & CAM-B3LYP & 20.\textbf{33} & 20.9\textbf{0} & 20.92 & 20.92 \\
        N$_2$, ${}^1\Pi_u$ & PBE & 1\textbf{2.64} & 17.\textbf{61} & 17.82 & 17.82 \\
        & B3LYP & 1\textbf{2.72} & 1\textbf{8.89} & 19.1\textbf{5} & 19.16 \\
        & CAM-B3LYP & \textbf{17.05} & 24.\textbf{62} & 24.90 & 24.90 \\
    \end{tabular}
\end{table}

\subsection*{Code}

The code used to obtain the results in this study can be found at the github repository \texttt{alieberherr/OTdensities}.

\bibliography{ot}